\documentclass[11pt,twoside]{article}

\usepackage{asp2006}
\usepackage{graphicx}

\begin{document}
\title{Effects of retro-lensing light curves near a black hole}   %%% Fill in title
\author{V. Karas}   %%% Fill in author names
\affil{Astronomical Institute, Academy of Sciences, Prague, Czech Republic}    %%% Fill in author affiliations

\begin{abstract} %%% Abstract to run on from here.
We model the light-curves from radiation-driven clouds near an accreting 
black hole. Taking into account the multiple images due to strong
gravitational lensing, we find that sharp spikes can significantly
enhance the observed flux. Following our previous work (Hor\'ak \& Karas
2006a,b) we assume that scattering of ambient light takes place in a
cloud that is in radial motion under a combined influence of black hole
gravity and the radiation field. The retro-lensed photons give rise to
peaks in the observed signal that follow with a characteristic time lag
after the direct-image photons. Duration of these features is very short
and the predicted polarization  varies abruptly on the time-scale
comparable with the light-crossing  time of the system -- a signature of
the photon orbit. We also  consider the polarization properties of
scattered light.
\end{abstract}

%*****************************************************************
\section{Introduction}
%***************************************************************** In
Holz \& Wheeler (2002) proposed that the electromagnetic signal from a
source located {\em in front of} a black hole may be significantly
enhanced by the influence of strong gravitational lensing. They called
this to be a retro-lensing effect (see also De Paolis et al.\ 2004), 
and considered it in the context of a
putative detection in the Solar system: a stellar-mass black hole
illuminated by light from the Sun and returning the ring of light back
towards the observer. Albeit a speculative set-up of the system, the
effect can find its application also in other sources containing a
radiation source close to the black hole horizon at (almost) perfect 
alignment with the direction towards observer. 

Related problems of strong-gravity lensing have been investigated by
various authors (e.g.\ Ohanian 1987; Virbhadra \& Ellis 2000; Bozza
2002). Retro-lensing images are formed by photon rays that make 180
degrees turn just above the black hole photon circular orbit. Although
the signal is usually weak in these images, favourable geometrical
arrangements are possible. Even more importantly, photons of these
higher-order images experience a characteristic time delay (see Bozza \&
Mancini 2004; \v{C}ade\'z \& Kosti\'c 2005), and this could help to
reveal the signatures of  strong gravitational field in the system which
may show up due to light signal making a complete turn around the black
hole (Bursa et al. 2007; Fukumura et al.\ 2008, 2009).

Here we would like to stress a point, originally proposed in Hor\'ak \&
Karas (2006a), that characterising time delays together with the help of
polarization information could constrain the model parameters to greater
confidence than what is possible with only photometric light-curves.
Furthermore, if the source of light is in a rapid fall towards the black
hole then the retro-lensing signal become {\em enhanced} with  respect
to the primary signal, thereby emphasising the importance of the
indirect images that are a specific kind of signature of the a black
hole horizon.

In fact, all higher-order images suffer from the attenuating influence
of the light bending that reduces their luminosity, unless a special
geometrical alignment of the source and the observer occurs and favours
the opposite effect of magnification in a caustic. This can be
quantified by the gain factor which determines the ratio of fluxes
received in retro-lensed and direct images.

%*****************************************************************
\section{Light intensity and polarization}
%***************************************************************** In
Hor\'ak \& Karas (2006b) we considered a cloud of particles moving
through the radiation field of a standard thin accretion disc. Primary
photons from the disc are scattered by electrons in the cloud near the
symmetry axis, they are beamed in the direction of the cloud motion and
polarized by Thomson scattering.

The electron distribution is considered isotropic in the cloud comoving
frame. We derived simple formulae for frequency-integrated Stokes 
parameters $I$, $Q$ and $U$ of the scattered radiation (Hor\'ak \& Karas
2006a,b): $I=A[(1+\mathcal{A})(T^{tt}+T^{ZZ})+
\mathcal{B}(T^{tt}-3T^{ZZ})-2\mathcal{A}T^{tZ}]$,
$Q=A\left(T^{YY}-T^{XX}\right)$, $U=-2\,A\,T^{XY},$ where
$\mathcal{A}\equiv\textstyle{\frac{4}{3}}\;\langle\gamma_\mathrm{e}^2\beta_\mathrm{e}^2\rangle$,
$\mathcal{B}\equiv 1-\bar{\sigma}$, where
$\bar{\sigma}\equiv\left<\beta_\mathrm{e}^{-1}\gamma_\mathrm{e}^{-2}\ln[\gamma_\mathrm{e}(1+\beta_\mathrm{e})]\right>$;
$\beta_\mathrm{e}$,  $\gamma_\mathrm{e}$ are velocity and the Lorentz
factor corresponding to an  individual electron, while the angle
brackets denote the averaging over the particle distribution in the
cloud comoving frame (see Hor\'ak \& Karas 2006b for notation and
details of the calculation).

The Stokes parameters are evaluated in the polarization frame comoving
with the cloud (one basis vector is pointed along the direction of the
scattered radiation, while the other two basis vectors are perpendicular
to it and to each other). The incident unpolarized radiation comes into
the formulae as components of the stress-energy tensor
$T^{\alpha\beta}$. The resulting polarization  is linear, and so the
fourth Stokes parameter $V$ vanishes.

%*****************************************************************
%\section{Motion of the cloud}
%***************************************************************** The
Total four-force $f^\alpha$ acting on the cloud is a superposition of 
the radiation and inertial terms. The cloud motion is solved in the 
spacetime of Schwarzschild black hole (radius $r_\mathrm{s}$). The
radiation field influences the bulk motion of the cloud as well as the 
local electron distribution in the cloud frame.  We find two critical
velocities at which the polarization vector changes its orientation
between transversal  and longitudinal one.

    \begin{table}[tbh]
      \begin{center}
        \begin{tabular}{cc}
           \includegraphics[width=0.44\textwidth]{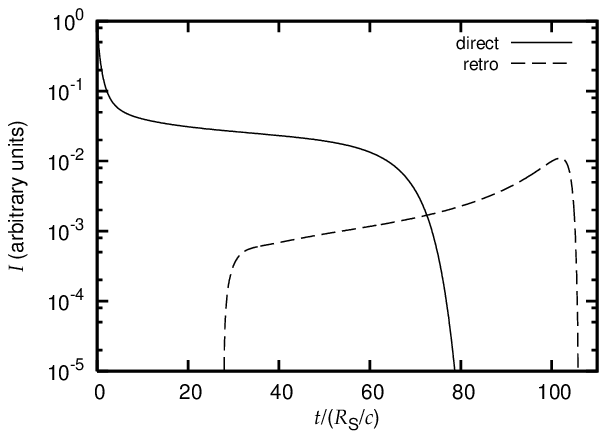}
            &
           \includegraphics[width=0.44\textwidth]{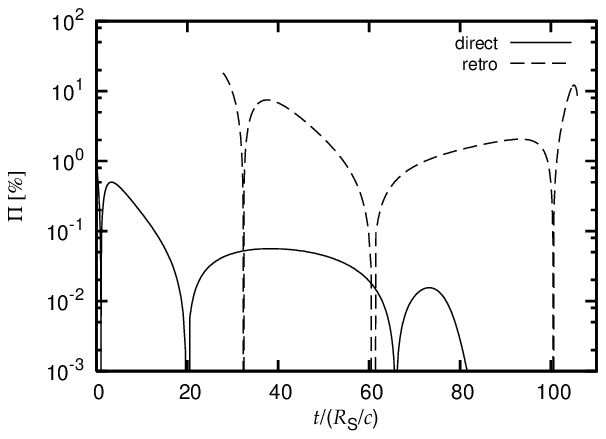}
           \\          
           \includegraphics[width=0.44\textwidth]{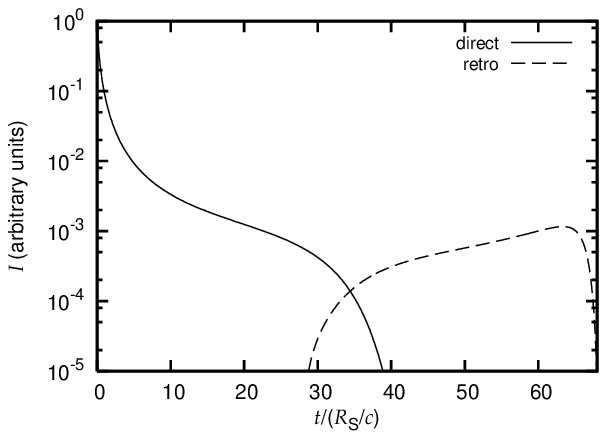}
           &
           \includegraphics[width=0.44\textwidth]{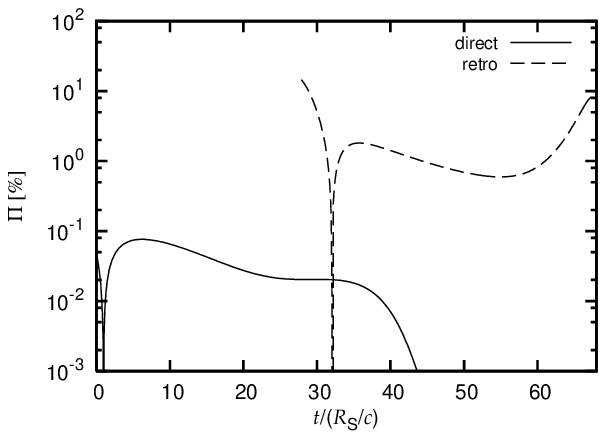}
        \end{tabular}
      \end{center}
  Fig.~1--Comparison of two typical cases with the identical initial
  conditions except for the cloud temperature: a cold cloud (upper
  panels, versus a warm cloud (lower panels, the electron Lorentz factor
  is 3 at start). Examples are shown of intensity (left panels) and
  polarization (middle panels) light-curves. Contributions of the
  retro-lensing images have been summed together (dashed line); they are
  clearly distinguished from the signal produced by the direct-image
  photons (solid line). Polarization vanishes at the moment when  the
  cloud crosses one of the curves $\beta_1(\xi)$, $\beta_2(\xi)$.   The
  view angle was $i=5$~deg in both cases  (figure adapted from Hor\'ak
  \& Karas 2006b).
    \end{table}

%*****************************************************************
%\section{Retro-lensing light-curves and polarization}
%***************************************************************** When
determining the temporal evolution of observed intensity and
polarization we consider the first three images of the observed
radiation -- the direct one and two retro-lensed images. The latter are
formed by rays making a turn around the black hole. For small
inclination angles these images take the form of Einstein arcs.

The retro-lensed photons give rise to peaks in the observed signal
occurring  with a characteristic mutual time lag after the direct-image
photons. Duration of these features is very short and comparable to the
light crossing time. They typically contribute about 10 percent of the
scattered flux at most, but  this result can be quite different for
matter infalling on to the star. In that case the scattered photons are
boosted in the downward direction and a considerable amount of light is
directed on to the photon orbit. As a result, the retro-lensed images
become much more pronounced and they cause a brief flash of light, as
illustrated in  figure~1.

%*****************************************************************
\section*{Conclusions}
%*****************************************************************
Although we described here a very simplistic model, our calculation is
self-consistent in the sense that the motion of the gaseous blob and of
the illuminating as well as scattered photons can be mutually
interconnected. This interaction gives rise to the resulting lightcurves
and also the predicted polarization at the observer. 

We concentrated ourselves on gravitational effects and 
compared the flux intensities and the polarization magnitudes
of direct and retro-lensing images. We noticed the mutual delay between
the peaks of the observed signal, formed by photons of different orders. 
The detected time delays are expected to be approximately the light circle 
time near the photon orbit. This value is characteristic to the effect 
and it is proportional to the black hole mass.

\medskip%\acknowledgements %%% Text of acknowledgements runs on after this command.

We acknowledge the Czech-US collaboration project No.~ME09036.

%*****************************************************************

\end{document}